\documentclass[amstex,12pt]{article}
\usepackage{amssymb}
\usepackage{amsmath}
\usepackage{IEEEtrantools}
\usepackage{pdflscape}
\usepackage{afterpage}
\usepackage{capt-of}

\usepackage{graphicx}
\usepackage{epstopdf}
\usepackage{subfig}

\usepackage{changepage}

\numberwithin{equation}{section}

\newcommand{\be}{\begin{equation}}
\newcommand{\ee}{\end{equation}}
\newcommand{\bea}{\begin{eqnarray}}
\newcommand{\eea}{\end{eqnarray}}
\newcommand{\non}{\nonumber}
\newcommand{\id}{\mathbb{I}}

\newcommand{\tr}{\mathop{\rm tr}\nolimits}
\newcommand{\diag}{\mathop{\rm diag}\nolimits}

\usepackage[usenames,dvipsnames]{color}

\begin{document}

\begin{titlepage}
\strut\hfill UMTG--295
\vspace{.5in}
\begin{center}

\LARGE New $D_{n+1}^{(2)}$ K-matrices with quantum group symmetry\\
\vspace{1in}
\large 
Rafael I. Nepomechie \footnote{Physics Department,
P.O. Box 248046, University of Miami, Coral Gables, FL 33124 USA, nepomechie@miami.edu}
and Rodrigo A. Pimenta  \footnote{Instituto de F\'{i}sica de S\~{a}o Carlos, Universidade de S\~{a}o Paulo, Caixa
Postal 369, 13566-590, S\~{a}o Carlos, SP, Brazil, pimenta@ifsc.usp.br}
\\[0.8in]
\end{center}

\vspace{.5in}

\begin{abstract}
    We propose new families of solutions of the $D_{n+1}^{(2)}$ boundary Yang-Baxter 
    equation. The open spin-chain transfer matrices constructed with these 
    K-matrices have quantum group symmetry corresponding to removing one node from
    the $D_{n+1}^{(2)}$ Dynkin diagram, namely, $U_{q}(B_{n-p}) 
    \otimes U_{q}(B_{p})$, where $p=0, \ldots, n$. These transfer 
    matrices also have a $p \leftrightarrow n-p$ duality symmetry. 
    These symmetries help to account for the degeneracies in the 
    spectrum of the transfer matrix.
\end{abstract}

\end{titlepage}

\setcounter{footnote}{0}

\section{Introduction}\label{sec:intro}

An important class of solutions of the Yang-Baxter equation is given
by the so-called trigonometric (or hyperbolic) R-matrices
\cite{Jimbo:1985ua, Bazhanov:1986mu}, which are associated with affine
Lie algebras.  These R-matrices, which are basic building blocks of
integrable quantum spin chains, were derived using quantum 
groups.\footnote{R-matrices have recently also been derived from 
four-dimensional gauge theory \cite{Costello:2017dso, Costello:2018gyb}.}
Ironically, the periodic local integrable quantum spin chains of finite
length constructed with these R-matrices do not exhibit any quantum
group symmetry, because the periodic boundary conditions are not
compatible with such symmetry.\footnote{Periodic spin chains can have 
quantum group symmetry if one allows non-local interactions 
\cite{Karowski:1993nw}.}  
However, quantum group symmetry can be
realized in {\em open} local integrable quantum spin chains of finite
length, provided that the boundary conditions are suitably chosen.
Integrable boundary conditions are provided by solutions of the
boundary Yang-Baxter equation \cite{Cherednik:1985vs}, which are often
called K-matrices or reflection matrices.

For trigonometric R-matrices and corresponding K-matrices
\cite{Batchelor:1996np, Malara:2004bi} associated with several series of affine Lie
algebras $\hat g$ (namely, $A_{2n}^{(2)}$, $A_{2n-1}^{(2)}$,
$B_{n}^{(1)}$, $C_{n}^{(1)}$ and $D_{n}^{(1)}$), the corresponding
open spin-chain transfer matrices \cite{Sklyanin:1988yz,
Mezincescu:1990uf} have recently been shown to have quantum group
symmetry corresponding to removing one node from the $\hat g$ Dynkin
diagram \cite{Nepomechie:2018dsn}. (The $A_{n-1}^{(1)}$ case was 
discussed in \cite{Doikou:1998ek}).  
However, the $D_{n+1}^{(2)}$ case
was not considered in \cite{Nepomechie:2018dsn}, since most of the
needed K-matrices were not known.\footnote{The $D_{n+1}^{(2)}$
R-matrix is -- by far -- the most complicated of the R-matrices in
\cite{Jimbo:1985ua, Bazhanov:1986mu}, which may explain why few of the
corresponding K-matrices had heretofore been found. Due to the complexity of the R-matrix, we do not prove that the proposed K-matrices satisfy the boundary Yang-Baxter equation (\ref{BYBEm}) and the duality property (\ref{Rmatprop}), but only check these results for small values of $n$.} The principal aim
of this note is to find the necessary K-matrices, and then to extend
the program in \cite{Nepomechie:2018dsn} to the $D_{n+1}^{(2)}$ case
($n=1, 2, \ldots$).  We show that the open spin-chain transfer
matrices constructed with these K-matrices indeed have quantum group
symmetry corresponding to removing the $p^{th}$ node from the
$D_{n+1}^{(2)}$ Dynkin diagram, namely, $U_{q}(B_{n-p}) \otimes
U_{q}(B_{p})$, where $p=0, \ldots, n$.\footnote{The K-matrices for the
case $p=n$ were already known \cite{Martins:2000xie, Malara:2004bi},
and the $U_{q}(B_{n})$ symmetry for this case was noted in
\cite{Nepomechie:2017hgw}.} A special feature of the $D_{n+1}^{(2)}$
case, which is present also for $C_{n}^{(1)}$ and $D_{n}^{(1)}$ cases
\cite{Nepomechie:2018dsn}, is the existence of an additional $p
\leftrightarrow n-p$ duality symmetry.

The outline of this paper is as follows.  The new K-matrices are
presented in Sec.  \ref{sec:Kmat}.  The construction of the open-chain
transfer matrix is briefly reviewed in Sec. \ref{sec:transfer}.  The
quantum group symmetry of the transfer matrix is demonstrated in
Sec. \ref{sec:QG}, and the duality symmetry of the transfer matrix is shown
in Sec. \ref{sec:duality}.  These symmetries are used in Sec.
\ref{sec:degen} to explain the degeneracies in the spectrum of the
transfer matrix for generic values of the anisotropy parameter $\eta$.
Sec. \ref{sec:conclusion} contains a brief conclusion. Since the $D_{n+1}^{(2)}$ case 
shares many similarities with other cases analyzed in \cite{Nepomechie:2018dsn}, we 
try to avoid repeating here previous results as much as possible, and 
instead emphasize the results that are different.

\section{R and K matrices}\label{sec:Kmat}

Let $R(u)$ denote the $D_{n+1}^{(2)}$ R-matrix \cite{Jimbo:1985ua}, 
following the conventions in \cite{Nepomechie:2017hgw}. This is a
$(2n+2)^{2} \times (2n+2)^{2}$ matrix, which is a function of the 
spectral parameter $u$ and the anisotropy parameter $\eta$,
and satisfies 
the Yang-Baxter equation
\be
R_{12}(u - v)\,  R_{13}(u)\, R_{23}(v) = R_{23}(v)\,  R_{13}(u)\, R_{12}(u - v)
\,.  \label{YBE}
\ee
We also use the following additional properties of this R-matrix: $PT$ symmetry
\be
R_{21}(u) \equiv {\cal P}_{12}\, R_{12}(u)\, {\cal P}_{12} 
= R_{12}^{t_1 t_2}(u) \,,
\label{PT}
\ee
unitarity
\be
R_{12}(u)\ R_{21}(-u) = \zeta(u)\, \id\otimes\id  \,,
\label{unitarity}
\ee
where $\zeta(u)$ is given by
\be
\zeta(u) =\xi(u)\, \xi(-u)\,, \qquad 
\xi(u)=4\sinh(u +2\eta) \sinh(u +2 n \eta) \,,
\label{xi}	    
\ee
and crossing symmetry
\be
R_{12}(u)=V_1\, R_{12}^{t_2}(-u-\rho)\, V_1
= V_2^{t_2}\, R_{12}^{t_1}(-u-\rho)\, V_2^{t_2} \,, \qquad 
\rho = -2 n\eta \,,
\label{crossing}
\ee
where the crossing matrix $V$ can be found in 
\cite{Nepomechie:2017hgw}.

Let $K^{R}(u)$ denote the ``right'' $D_{n+1}^{(2)}$ K-matrix, which 
is a $(2n+2) \times (2n+2)$ matrix that satisfies
the boundary Yang-Baxter equation \cite{Cherednik:1985vs, Sklyanin:1988yz, Ghoshal:1993tm}
\be
R_{12}(u - v)\, K^{R}_1(u)\ R_{21} (u + v)\, K^{R}_2(v)
= K^{R}_2(v)\, R_{12}(u + v)\, K^{R}_1(u)\, R_{21}(u - v)  \,.
\label{BYBEm}
\ee
In view of the expected $U_{q}(B_{n-p}) \otimes U_{q}(B_{p})$ 
symmetry, we look for solutions of the following block-diagonal form
{\small
\be 
K^{R}(u) = K^{R}(u,p) = 
\left( \begin{array}{c|c|cc|c|c}
k_{-}(u)\, \id_{p \times p} & & & & & \\
\hline
& g(u)\, \id_{(n-p) \times (n-p)} & & & &\\
\hline
& & k_{1}(u) & k_{2}(u) & &\\
& & k_{2}(u) & k_{1}(u) & & \\
\hline
& & & & g(u)\, \id_{(n-p) \times (n-p)} &  \\
\hline
& & & & & k_{+}(u)\, \id_{p \times p}
\end{array} \right) \,,
\label{KR}
\ee}
where $p=0, \ldots, n$ and $n = 1, 2, \ldots$.
We impose the regularity constraint $K^{R}(0,p)=\id$, and proceed to 
solve for the unknown functions following the method in the appendix 
of \cite{Mezincescu:1990ui}, for small values of $n$. The following 
pattern emerges
\begin{align}
k_{\mp}(u) &= e^{\mp 2u} \,, \non \\
g(u) &= \frac{\cosh(u-(n-2p)\eta + 
\frac{i\pi}{2}\varepsilon)}{\cosh(u+(n-2p)\eta - \frac{i\pi}{2}\varepsilon)} \,, \non \\
k_{1}(u)  &= \frac{\cosh(u) \cosh((n-2p)\eta + \frac{i\pi}{2}\varepsilon)}
{\cosh(u+(n-2p)\eta)+\frac{i\pi}{2}\varepsilon)} 
\,, \non \\
k_{2}(u)  &= -\frac{\sinh(u) \sinh((n-2p)\eta + \frac{i\pi}{2}\varepsilon)}
{\cosh(u+(n-2p)\eta + \frac{i\pi}{2}\varepsilon)} 
\,,
\label{functions}
\end{align}
where $\varepsilon$ can take either of two values: $\varepsilon=0$ 
or  $\varepsilon=1$. Unless otherwise 
noted, all the results in this paper hold for both values of $\varepsilon$.
We have explicitly verified all of the solutions 
(\ref{KR}), (\ref{functions}) up to $n=10$, and we conjecture 
that they are valid for all $n$.

Note that the solutions with
$p=\frac{n}{2}$ ($n$ even) and $\varepsilon=0$ are diagonal; 
otherwise, the solutions are non-diagonal.
The solutions with $p=n$, as well as the diagonal
solution with $n=2, p=1, \varepsilon=0$, were previously known 
\cite{Martins:2000xie, Malara:2004bi}; to our knowledge, all other 
solutions are new.

In order to maximize the symmetry of the spin chain, we impose the 
``same'' boundary conditions on the two ends, which corresponds to 
taking the ``left'' K-matrix $K^{L}(u)$ to be
\be
K^{L}(u) = K^{L}(u,p) = K^{R}(-u-\rho,p)\, M\,,
\label{KL}
\ee
where $M$ is a diagonal matrix defined in terms of the crossing matrix $V$ by 
$M = V^{t}\, V $, whose explicit expression can be found in 
\cite{Nepomechie:2017hgw}.

\section{Transfer matrix}\label{sec:transfer}

Using the R-matrix and the K-matrices (\ref{KR}) and (\ref{KL}), we can 
construct the transfer matrix of an integrable open spin chain of 
length $N$ \cite{Sklyanin:1988yz}
\be
t(u,p) = \tr_a K^{L}_{a}(u,p)\, T_a(u)\,  K^{R}_{a}(u,p)\, \widehat{T}_a(u) \,, 
\label{transfer}
\ee
where the single-row monodromy matrices are defined by
\begin{align} 
T_a(u) &= R_{aN}(u)\ R_{a N-1}(u)\ \cdots R_{a1}(u) \,,  \non \\
\widehat{T}_a(u) &= R_{1a}(u)\ \cdots R_{N-1 a}(u)\ R_{Na}(u) \,,  
\label{monodromy}
\end{align}
and the trace in (\ref{transfer}) is over the ``auxiliary'' space, which 
is denoted by $a$. The transfer matrix satisfies the 
fundamental commutativity property
\be
\left[ t(u,p) \,, t(v,p) \right] = 0 \hbox{   for all   } u \,, v \,,
\label{commutativity} 
\ee
and contains the Hamiltonian ${\cal H}(p) \sim t'(0,p)$
as well as higher local conserved quantities.

\section{Quantum group symmetry}\label{sec:QG}

We now argue that the transfer matrix (\ref{transfer}) has the 
quantum group symmetry $U_{q}(B_{n-p}) \otimes U_{q}(B_{p})$. The 
argument is very similar to the one in \cite{Nepomechie:2018dsn}. The 
key idea is to perform a $p$-dependent gauge transformation of the R and K matrices
\be
\tilde{R}_{12}(u,p) = B_{1}(u,p)\, R_{12}(u)\, B_{1}(-u,p) = 
B_{2}(-u,p)\, R_{12}(u)\, B_{2}(u,p)\,,
\label{gaugeR}
\ee
and 
\begin{align}
\tilde{K}^{R}(u,p) &= B(u,p)\, K^{R}(u,p)\, B(u,p)\,, \non \\
\tilde{K}^{L}(u,p) &= B(-u,p)\, K^{L}(u,p)\, B(-u,p)\,,
\label{gaugeK}
\end{align}
where $B(u,p)$ is the diagonal matrix 
\be
B(u,p) =  \diag \big( \underbrace{e^{u}\,, \ldots\,, 
e^{u}}_{p}\,, \underbrace{1\,, \ldots\,, 1}_{2n+2-2p}\,, 
\underbrace{e^{-u}\,, \ldots\,, 
e^{-u}}_{p}\big) \,.
\label{Bgauge}
\ee
Evidently, the gauge-transformed K-matrix $\tilde{K}^{R}(u,p)$ is 
the same as (\ref{KR}), except that the functions $k_{\mp}(u)$ are replaced by 1.
The gauge-transformed (single-row) monodromy matrix
\be
\tilde{T}_a(u,p) = \tilde{R}_{aN}(u,p)\ \tilde{R}_{a N-1}(u,p)\ 
\cdots \tilde{R}_{a1}(u,p)  
\ee
has the asymptotic behavior
\be
\tilde{T}_a(u,p) \sim e^{\pm 2 N u}\, \tilde{T}^{\pm}_a(p) \quad \mbox{  
for  } \quad  u \rightarrow \pm \infty \,,
\ee
where
\be
\tilde{T}^{\pm}_a(p) = \tilde{R}_{aN}^{\pm}(p)\ \tilde{R}_{a N-1}^{\pm}(p)\\ 
\cdots \tilde{R}_{a1}^{\pm}(p)\ 
\label{tildeTpm}
\ee
and
\be
\tilde{R}^{\pm}(p) = \lim_{u\rightarrow \pm \infty} e^{\mp 2u} 
\tilde{R}(u,p) \,.
\label{Rtildeasym}
\ee
The important point is that the operators
$\tilde{T}^{\pm}_{i,j}(p) = \left(\tilde{T}^{\pm}_a(p)\right)_{ij}$ can be expressed in terms of (the quantum enveloping 
algebra of) the unbroken $D_{n+1}^{(2)}$ generators, i.e. the
generators of $U_{q}(B_{n-p}) \otimes U_{q}(B_{p})$, see Appendix 
\ref{sec:QGgens}. Hence, in order to demonstrate the 
quantum group symmetry of the transfer matrix, it suffices to show 
that
\be
\left[  \tilde{T}^{\pm}_{i,j}(p)\,, t(u,p)\right] = 0 \qquad i, j = 
1, 2, \ldots, 2n+2
\,.
\label{qgsymmetry}
\ee
The proof of (\ref{qgsymmetry}) in \cite{Nepomechie:2018dsn} carries 
over readily to the $D_{n+1}^{(2)}$ case, except for Lemma 1
\be
\left[\tilde{R}_{12}^{\pm}(p)\,, \tilde{K}^{R}_{2}(u,p)
\right] = 0 \,.
\label{lemma1}
\ee
We have verified this relation explicitly for small values of $n$ 
(for both $\varepsilon = 0$ and $\varepsilon = 1$), 
and we conjecture that it is true for all $n$.

The $(2n+2)$-dimensional vector space at each site decomposes under 
$U_{q}(B_{n-p}) \otimes U_{q}(B_{p})$ simply as the direct sum of vector 
representations of each factor, i.e.
\be
(2(n-p)+1,1) \oplus (1, 2p+1) \,.
\label{rep}
\ee

\section{Duality symmetry}\label{sec:duality}

The transfer matrix (\ref{transfer}) also has the $p \leftrightarrow 
n-p$ duality symmetry
\be
{\cal U}\, t(u,p)\, {\cal U}^{-1} = f(u,p)\, t(u,n-p)  \,, 
\label{duality}
\ee 
where ${\cal U}$ is the quantum-space operator 
\be
{\cal U} = U_{1}\ldots U_{N} \,,
\label{Ucal}
\ee
$U$ is the block matrix
\begin{align}
U &= \begin{cases}
\left(\begin{array}{c|cc|c}
&  & & \id_{n \times n} \\
\hline
& 1 & 0 &\\
& 0 & -1 & \\
\hline
-\id_{n \times n}  & & & \\ 
\end{array}\right)_{(2n+2) \times (2n+2)} 
& \mbox{for $n$ even} \,, \\ \\
\left(\begin{array}{c|cc|c}
&  & & \id_{n \times n} \\
\hline
& 0 & 1 &\\
& -1 & 0 & \\
\hline
-\id_{n \times n}  & & & \\ 
\end{array}\right)_{(2n+2) \times (2n+2)} 
& \mbox{for $n$ odd } \,,
\end{cases}
\label{Umat}
\end{align}    
which satisfies $U\, U^{t} = \id$,
and $f(u,p)$ is a scalar function given below (\ref{ffunction}). The proof of 
(\ref{duality}) is similar 
to the one in \cite{Nepomechie:2018dsn}. It makes use of the following 
properties of the $D_{n+1}^{(2)}$ R-matrix
\begin{align}
U_{1}\, R_{12}(u)\, U_{1}^{t} &= W_{2}^{t}(u)\, R_{12}(u)
\left(W_{2}^{t}(u)\right)^{-1} \,, \non \\
U_{2}\, R_{12}(u)\, U_{2}^{t} &= \left(W_{1}(u)\right)^{-1} R_{12}(u)\, 
W_{1}(u) \,,
\label{Rmatprop}
\end{align}
where $W(u)$ is the block matrix
\begin{align}
W(u) &= \begin{cases}
\left(\begin{array}{c|cc|c}
&  & & -e^{-u}\id_{n \times n} \\
\hline
& -1 & 0 &\\
& 0 & 1 & \\
\hline
e^{u}\id_{n \times n}  & & & \\ 
\end{array}\right)_{(2n+2) \times (2n+2)} 
& \mbox{for $n$ even} \,, \\ \\
\left(\begin{array}{c|cc|c}
&  & & -e^{-u}\id_{n \times n} \\
\hline
& 0 & -1 &\\
& 1 & 0 & \\
\hline
e^{u}\id_{n \times n}  & & & \\ 
\end{array}\right)_{(2n+2) \times (2n+2)} 
& \mbox{for $n$ odd } \,.
\end{cases}
\label{Wmat}
\end{align}   
We have verified the properties (\ref{Rmatprop}) for small values of $n$, and we 
conjecture that they are true for all $n$.
Moreover, the K-matrices (\ref{KR}) and (\ref{KL}) satisfy
\begin{align}
W(u)\, K^{R}(u,p)\, W^{t}(u) &= f^{R}(u,p)\,  K^{R}(u,n-p) \,, \non \\
\left(W^{t}(u)\right)^{-1} K^{L}(u,p) \left(W(u)\right)^{-1} &= f^{L}(u,p)\,  K^{L}(u,n-p) \,,
\label{dualityK}
\end{align}
where $f^{R}(u,p)$ and $f^{L}(u,p)$ are scalar functions given by 
\begin{align}
f^{R}(u,p) &= \frac{\cosh(u-(n-2p)\eta + \frac{i\pi}{2}\varepsilon)}
{\cosh(u+(n-2p)\eta - \frac{i\pi}{2}\varepsilon)} \,,  \non \\
f^{L}(u,p) &= 
\frac{\cosh(u-(n+2p)\eta + \frac{i\pi}{2}\varepsilon)}
{\cosh(u-(3n-2p)\eta - \frac{i\pi}{2}\varepsilon)}
\,.
\end{align}
Using the properties (\ref{Rmatprop}) and (\ref{dualityK}), it is 
now straightforward to show \cite{Nepomechie:2018dsn} that the transfer 
matrix has the duality symmetry (\ref{duality}), where $f(u,p)$ is 
given by
\be
f(u,p) = f^{L}(u,p)\, f^{R}(u,p) \,.
\label{ffunction}
\ee 
Consequently, for each eigenvalue 
$\Lambda(u,p)$ of $t(u,p)$, there is a corresponding 
eigenvalue $\Lambda(u,n-p)$ of $t(u,n-p)$ such that 
\be
\Lambda(u,p)  = f(u,p)\, \Lambda(u,n-p) \,.
\label{duality2}
\ee

\subsection{Action of duality on the quantum group}

Under a duality transformation, the operators 
$\tilde{T}^{\pm}_{i,j}(p)$ (\ref{tildeTpm}) transform as follows
\be
{\cal U}\, \tilde{T}^{\pm}_{a}(p)\, {\cal U}^{-1} = U_{a}^{-1}\, \tilde{T}^{\pm}_{a}(n-p)\, 
U_{a} \,.
\label{prop3}
\ee
The proof is similar to the one in \cite{Nepomechie:2018dsn}. In 
particular, the generators (\ref{leftgensvectorrep}), 
(\ref{rightgensvectorrep}) transform as\footnote{The relations 
(\ref{dualitygenerators}) are consistent by virtue of the identities
\begin{align}
U^{2}\, H_{i}^{(l)}(p) &=  H_{i}^{(l)}(p)\, U^{2}\,,    &
U^{2}\, E_{i}^{\pm\, (l)}(p) &= \nu_{i}(p)\, E_{i}^{\pm\, (l)}(p)\, 
U^{2}\,,  \non \\
U^{2}\, H_{i}^{(r)}(p) &=  H_{i}^{(r)}(p)\, U^{2}\,,    &
U^{2}\, E_{i}^{\pm\, (r)}(p) &= \nu_{i}(n-p)\, E_{i}^{\pm\, (r)}(p)\, 
U^{2}\,. \non
\end{align}}
\begin{align}
U\, H_{i}^{(l)}(p)\, U^{-1} &=   H_{i}^{(r)}(n-p)\,, &
U\, E_{i}^{\pm\, (l)}(p)\, U^{-1} &=   \nu_{i}(p)\, E_{i}^{\pm\, (r)}(n-p)\,,
\qquad i = 1, 2, \ldots, n-p \,, \non \\
U\, H_{i}^{(r)}(p)\, U^{-1} &=   H_{i}^{(l)}(n-p)\,, &
U\, E_{i}^{\pm\, (r)}(p)\, U^{-1} &=   E_{i}^{\pm\, (l)}(n-p)\,,
\qquad i = 1, 2, \ldots, p \,,
\label{dualitygenerators}
\end{align}
where 
\be
\nu_{i}(p) =
\begin{cases}
    -1 & \mbox{ if  $n=$ even and $i=n-p$} \\
    +1 & \mbox{otherwise}
    \end{cases} \,,
\ee    
and similarly for the coproducts.

\subsection{Self-duality}
For $p=\frac{n}{2}$ with $n$ even, the duality relation 
(\ref{duality}) implies that the transfer matrix is self-dual
\be
\left[ {\cal U}\,,  t(u, \tfrac{n}{2}) \right] = 0 \,,
\label{selfdual}
\ee
since $f(u,\frac{n}{2}) = 1$. This symmetry 
maps the representations $(1, {\bf R})$ and $({\bf R}, 1)$ (i.e., 
with ``left'' and ``right'' singlets, respectively) into each other;
and therefore these states are degenerate (i.e., have the same 
transfer-matrix eigenvalue). 

\subsubsection{Bonus symmetry for $\varepsilon=1$}\label{sec:bonus} 

For $p=\frac{n}{2}$ ($n$ even) and $\varepsilon=1$, there is an additional (``bonus'') 
symmetry, which leads to even higher degeneracies for the 
transfer-matrix eigenvalues. A similar phenomenon occurs for 
$C_{n}^{(1)}$ and $D_{n}^{(1)}$ \cite{Nepomechie:2018dsn}. Indeed, 
one can show in a similar way that the transfer matrix obeys
\be
    \left[ {\cal D}\,,  t(u, \tfrac{n}{2}) \right] = 0 \,,
\label{calDsymmetry}
\ee
where ${\cal D}$ is the quantum-space operator given by
\be
{\cal D} = D_{1} = D \otimes \id^{\otimes (N-1)}  \,,
\label{Dop}
\ee 
and $D$ is the ($u$-independent) matrix given by the 
gauge-transformed K-matrix
\be
D = \tilde{K}^{R}(u,\tfrac{n}{2}) \,,
\label{tildeKspecial}
\ee
which here is not diagonal. 

A state $|\Lambda\rangle$ that is a simultaneous eigenstate of $t(u, 
\tfrac{n}{2})$ and ${\cal U}$ (recall (\ref{selfdual})) is {\em not} 
necessarily an eigenstate of ${\cal D}$, since ${\cal U}$ and ${\cal 
D}$ do not commute. In such case, $|\Lambda\rangle$ and ${\cal D} |\Lambda\rangle$
are linearly independent eigenstates with the same transfer-matrix 
eigenvalue (recall (\ref{calDsymmetry})). In fact, the degeneracy of 
this eigenvalue becomes {\em doubled} as a consequence of the bonus 
symmetry \cite{Nepomechie:2018dsn}.

\section{Degeneracies of the transfer-matrix spectrum}\label{sec:degen}

For generic values of the anisotropy parameter $\eta$, the degeneracies 
in the spectrum of the transfer matrix (\ref{transfer}) mostly match with the predictions 
from the $U_{q}(B_{n-p}) \otimes U_{q}(B_{p})$ symmetry.
Exceptions include when $n$ is even and $p=\frac{n}{2}$ (in which case there is a 
self-duality symmetry (\ref{selfdual}); and, if $\varepsilon = 1$, 
there is also a bonus symmetry (\ref{calDsymmetry})) or when $n$ is odd and 
$p=\frac{n\pm 1}{2}$. Moreover, 
the spectrum exhibits a $p \rightarrow n-p$ duality symmetry.
We now consider some simple examples.

\subsection{Example 1: even $n$}\label{sec:neven}

We first consider the 
case $n=4$ (i.e., $D_{5}^{(2)}$) and $N=2$ (two sites). By direct diagonalization 
of the transfer matrix $t(u,p)$ for generic numerical values of $u$ and 
$\eta$, we find that the degeneracies are as follows: 
\begin{align}
& p=0:  &  &\{1, 1, 9, 9, 36, 44 \}\non \\
& p=1:  &  &\{1, 1, 3, 5, 21, 21, 21, 27 \}\non \\
& p=2:  &  &\begin{cases}
\{1, 1, 20, 25, 25, 28 \} & \mbox{for } \varepsilon=0\\
\{2, 20, 28, 50 \} & \mbox{for } \varepsilon=1
\end{cases}
\non \\
& p=3:  &  &\{1, 1, 3, 5, 21, 21, 21, 27 \}\non \\
& p=4:  &  &\{1, 1, 9, 9, 36, 44 \} \,.
\label{numericaldegen1x}
\end{align}
In other words, for $p=0$, one eigenvalue is repeated 44 times, 
another eigenvalue is repeated 36 times, etc.; and similarly for other values of $p$.
The fact that the degeneracies are the same for $p$ and $n-p$ is a 
consequence of the duality symmetry (\ref{duality}), (\ref{duality2}).

On the other hand, the quantum group symmetry when $n=4$ 
is $U_{q}(B_{4-p}) \otimes U_{q}(B_{p})$, and the 10-dimensional representation at 
each site (\ref{rep})
is $(9-2p,1) \oplus (1,2p+1)$. For generic values of $\eta$, the 
quantum group representations are the same as for the corresponding classical groups.
Performing the tensor-product decompositions using LieART \cite{Feger:2012bs}, we obtain
\begin{align}
& p=0: B_{4} & (\mathbf{9} \oplus\mathbf{1})^{\otimes 2} &= 2 
(\mathbf{1}) \oplus 
2(\mathbf{9}) \oplus \mathbf{36} \oplus \mathbf{44} \non \\
& p=1: B_{3}\otimes B_{1} & ((\mathbf{7},\mathbf{1}) \oplus 
(\mathbf{1},\mathbf{3}))^{\otimes 2} &= 2(\mathbf{1}, \mathbf{1}) 
\oplus (\mathbf{1}, \mathbf{3}) \oplus (\mathbf{1}, \mathbf{5}) 
\oplus 2 (\mathbf{7}, \mathbf{3}) \oplus (\mathbf{21}, \mathbf{1}) 
\oplus (\mathbf{27}, \mathbf{1})  \non \\
& p=2: B_{2}\otimes B_{2} & ((\mathbf{5},\mathbf{1}) \oplus 
(\mathbf{1},\mathbf{5}))^{\otimes 2} &= 2(\mathbf{1}, \mathbf{1}) 
\oplus (\mathbf{10}, \mathbf{1}) \oplus (\mathbf{1}, \mathbf{10})  
\oplus 2 (\mathbf{5}, \mathbf{5}) \oplus (\mathbf{14}, \mathbf{1}) 
\oplus (\mathbf{1}, \mathbf{14})  \,.
\label{LieART1x}
\end{align}
There is no need to display the tensor-product decompositions for 
$p>2$ due to the symmetry $p \rightarrow n-p$.

Comparing the degeneracies (\ref{numericaldegen1x}) with the 
corresponding tensor-product decompositions (\ref{LieART1x}), we 
see that they match, except for $p=2$. For the latter case, the 
degeneracies are larger, due to the self-duality symmetry (\ref{selfdual}) 
for even $n$ and $p=\frac{n}{2}$, which here maps $(\mathbf{1}, 
\mathbf{10})$ to $(\mathbf{10}, \mathbf{1})$ (resulting 
in a 20-fold degeneracy), and also maps $(\mathbf{1}, 
\mathbf{14})$ to $(\mathbf{14}, \mathbf{1})$ (resulting 
in a 28-fold degeneracy). If $\varepsilon=1$, then the 
bonus symmetry (\ref{calDsymmetry}) implies that 
the two $ (\mathbf{5}, \mathbf{5})$ are degenerate (giving rise to a 
50-fold degeneracy), as well as the 
two $(\mathbf{1}, \mathbf{1})$ (resulting in a 2-fold degeneracy).

\subsection{Example 2: odd $n$}\label{sec:nodd}

As a second example, we consider the 
case $n=5$  (i.e., $D_{6}^{(2)}$) and $N=2$ (two sites). By direct diagonalization 
of the transfer matrix $t(u,p)$ for generic numerical values of $u$ and 
$\eta$, we find that the degeneracies are as follows: 
\begin{align}
& p=0:  &  &\{1,1, 11, 11, 55, 65 \}\non \\
& p=1:  &  &\{1, 1, 3, 5, 27, 27, 36, 44\} \non \\
& p=2:  &  &\{1, 1, 10, 27, 49, 56 \}\non \\
& p=3:  &  &\{1, 1, 10, 27, 49, 56 \}\non \\
& p=4:  &  &\{1, 1, 3, 5, 27, 27, 36, 44\} \non \\
& p=5:  &  &\{1,1, 11, 11, 55, 65 \} \,.
\label{numericaldegen1}
\end{align}
We see again that the degeneracies are the same for $p$ and $n-p$, as a 
consequence of the duality symmetry (\ref{duality}), (\ref{duality2}).

On the other hand, the symmetry when $n=5$ 
is $U_{q}(B_{5-p}) \otimes U_{q}(B_{p})$, and the 12-dimensional 
representation at each site (\ref{rep})
is $(11-2p,1) \oplus (1,2p+1)$. The tensor-product decompositions are as follows:
\begin{align}
& p=0: B_{5} & (\mathbf{11} \oplus \mathbf{1})^{\otimes 2} &= 
2(\mathbf{1}) \oplus 2(\mathbf{11}) \oplus 
\mathbf{55} \oplus \mathbf{65} \non \\
& p=1: B_{4}\otimes B_{1} & ((\mathbf{9},\mathbf{1}) \oplus 
(\mathbf{1},\mathbf{3}))^{\otimes 2} &= 2(\mathbf{1}, \mathbf{1}) 
\oplus (\mathbf{1}, \mathbf{3}) \oplus (\mathbf{1}, \mathbf{5}) 
\oplus 2(\mathbf{9}, \mathbf{3}) \oplus (\mathbf{36}, \mathbf{1}) 
\oplus (\mathbf{44}, \mathbf{1})  \non \\
& p=2: B_{3}\otimes B_{2} & ((\mathbf{7},\mathbf{1}) \oplus 
(\mathbf{1},\mathbf{5}))^{\otimes 2} &= 2(\mathbf{1}, \mathbf{1}) 
\oplus (\mathbf{1}, \mathbf{10}) 
\oplus 2 (\mathbf{7}, \mathbf{5}) \oplus (\mathbf{1}, \mathbf{14}) 
\oplus (\mathbf{21}, \mathbf{1}) \oplus (\mathbf{27}, \mathbf{1})  \,.
\label{LieART1}
\end{align}
Again, there is no need to display the tensor-product decompositions for 
$p>2$ due to the symmetry $p \rightarrow n-p$.

Comparing the degeneracies (\ref{numericaldegen1}) with the 
corresponding tensor-product decompositions (\ref{LieART1}), we 
see that they match, except for $p=2$. For the latter case, the 
degeneracies are larger: the $(\mathbf{1}, \mathbf{14})$ and one  
$(\mathbf{7}, \mathbf{5})$ are degenerate (resulting in a 49-fold 
degeneracy); and the $(\mathbf{21}, \mathbf{1})$  and the other 
$(\mathbf{7}, \mathbf{5})$ are degenerate (resulting in a 56-fold 
degeneracy). Similar degeneracies for odd $n$ and $p=\frac{n\pm 1}{2}$
also occur for $C_{n}^{(1)}$ and $D_{n}^{(1)}$ \cite{Nepomechie:2018dsn}.

\section{Conclusions}\label{sec:conclusion}

We have found new $D_{n+1}^{(2)}$ K-matrices $K^{R}(u,p)$ (\ref{KR}), 
(\ref{functions}), for which the corresponding transfer matrix 
(\ref{transfer}) has $U_{q}(B_{n-p}) \otimes U_{q}(B_{p})$ symmetry 
(\ref{qgsymmetry}), as well as the $p \leftrightarrow 
n-p$ duality symmetry (\ref{duality}). For the special case
$p=\frac{n}{2}$ ($n$ even), the transfer matrix has a self-duality symmetry 
(\ref{selfdual}), and an additional ``bonus'' symmetry 
(\ref{calDsymmetry}) if $\varepsilon=1$. These symmetries account for 
most of the degeneracies of the spectrum of the transfer matrix, as 
illustrated in the examples of Sec. \ref{sec:degen}. The 
exceptions include the unusual degeneracy that occurs for $p=\frac{n\pm 
1}{2}$ ($n$ odd), noted in Sec. \ref{sec:nodd},
which is due to ``mixing'' of representations of {\em unequal} dimensions.
We expect that such degeneracies,
which occur also for $C_{n}^{(1)}$ and 
$D_{n}^{(1)}$ \cite{Nepomechie:2018dsn},
can be attributed to some discrete symmetries, 
which remain to be understood.

The following picture emerges about the symmetries of an integrable
spin chain of length $N$ constructed with a trigonometric R-matrix associated with
an affine Lie algebra $\hat g$: In the limit $N \rightarrow \infty$, 
the spin chain has the infinite-dimensional $U_{q}(\hat g)$ symmetry, regardless of boundary 
conditions, which is exploited in the vertex operator formalism 
(see e.g. \cite{Foda:1991wt, Davies:1992sva, Jimbo:1992, Jimbo:1994}). However, for finite $N$, the 
symmetry algebra of the spin chain is necessarily a finite-dimensional 
subalgebra of $U_{q}(\hat g)$. Maximal subalgebras of $U_{q}(\hat g)$ can presumably be obtained by removing 
one node from its Dynkin diagram. The present and 
previous \cite{Nepomechie:2018dsn, Doikou:1998ek} work describe the 
boundary conditions and the corresponding
integrable open spin chains 
with such symmetries, for all non-exceptional $\hat g$. Other boundary conditions presumably  lead to 
smaller symmetries.

\section*{Acknowledgments}
We thank A. Lima-Santos and A. Retore for discussions.
This work was supported by the
S\~ao Paulo Research Foundation (FAPESP) and the University of Miami
under the SPRINT grant \#2016/50023-5.  Additional support was
provided by a Cooper fellowship (RN) and by FAPESP/CAPES grant \# 
2017/02987-8 (RP).  RN thanks IFSC-USP and P. Pearce for warm hospitality.

\appendix

\section{Quantum group generators}\label{sec:QGgens}

We present here explicit expressions for the $U_{q}(B_{n-p}) \otimes U_{q}(B_{p})$ generators, in 
terms of which the operators $\tilde{T}^{\pm}_{i,j}(p)$ 
(\ref{tildeTpm}) can be 
expressed. Following \cite{Nepomechie:2018dsn}, we denote the generators 
corresponding to the simple roots of the ``left'' algebra $g^{(l)} 
\equiv B_{n-p}$ and the ``right'' algebra $g^{(r)} \equiv B_{p}$ by
\be
H^{(l)}_{i}(p)\,, \quad E^{\pm\, (l)}_{i}(p) \,, \qquad i = 1, \ldots, 
n-p\,,  \non
\ee
and 
\be
H^{(r)}_{i}(p)\,, \quad E^{\pm\, (r)}_{i}(p) \,, \qquad i = 1, \ldots, p \,, \non
\ee
respectively.
The ``left'' generators satisfy the commutation relations
\begin{align}
\left[H^{(l)}_i(p)\,, H^{(l)}_j(p) \right] &= 0 \,, \non\\
\left[H^{(l)}_i(p)\,, E^{\pm\, (l)}_j(p) \right] &= \pm \alpha_i^{(j)}E^{\pm\, (l)}_j(p) \,, \non\\
\left[E^{+\, (l)}_i(p)\,, E^{-\, (l)}_j(p) \right] &=\delta_{i,j}\sum_{k=1}^{n-p}\alpha_k^{(j)}H^{(l)}_k(p) \,,
\label{leftcommutators}
\end{align}
and the ``right'' generators similarly satisfy the commutation relations 
\begin{align}
\left[H^{(r)}_i(p)\,, H^{(r)}_j(p) \right] &= 0 \,, \non\\
\left[H^{(r)}_i(p)\,, E^{\pm\, (r)}_j(p) \right] &= \pm 
\alpha_i^{(j)}E^{\pm\, (r)}_j(p) \,, \non\\
\left[E^{+\, (r)}_i(p)\,, E^{-\, (r)}_j(p) \right] 
&=\delta_{i,j}\sum_{k=1}^{p}\alpha_k^{(j)}H^{(r)}_k(p) \,.
\label{rightcommutators}
\end{align}
Moreover, the ``left'' and  ``right'' generators commute with each 
other. 
The simple roots $\{\alpha^{(1)}, 
\ldots,  \alpha^{(m)} \}$ of $B_{m}$  (where $m$ is either $n-p$ or $p$) in the orthogonal basis are given by
\begin{align}
\alpha^{(j)} &=   e_{j} - e_{j+1} \,, \qquad  j =  1,\ldots, m-1\,,   
\non \\
\alpha^{(m)} &= e_{m} \,,
\end{align}
where $e_{j}$ are the elementary $m$-dimensional basis vectors 
$(e_{j})_{i} = \delta_{i,j}$.

In terms of the $D_{n+1}^{(2)}$ generators
\begin{align}
    H_{i} &= e_{i,i} - e_{2n+3-i,2n+3-i} \,, \qquad i = 1, \ldots, n \,, \non \\
    E^{+}_{i} &= e_{i,i+1} + e_{2n+2-i, 2n+3-i} \,, \qquad i = 1, \ldots, n-1 \,, \non \\
    E^{+}_{n} &= \frac{1}{\sqrt{2}}\left(e_{n,n+1}+e_{n,n+2}-e_{n+2,n+3}-e_{n+1,n+3}\right) \,, \non \\
    E^{+}_{0} &= 
    \frac{1}{\sqrt{2}}(-1)^{n}\left(e_{n+1,1}-e_{n+2,1}+e_{2n+2,n+1}-e_{2n+2,n+2}\right)
    \,, \non \\
    E^{-}_{i} &= (E^{+}_{i})^{t} \,, \qquad\qquad\qquad i = 0, 1, 
    \ldots, n \,,
    \label{gensp0}
\end{align}
where $e_{ij}$ are the elementary $(2n+2) \times (2n+2)$ matrices, 
the ``left and ``right'' generators are given by
\be
H_{i}^{(l)}(p) = H_{p+i} \,,  \qquad
E^{\pm\, (l)}_{i}(p) =  E^{\pm}_{p+i} \,, \qquad i = 1, \ldots, n-p \,,
\label{leftgensvectorrep}
\ee
and
\be
H_{i}^{(r)}(p) = -H_{p+1-i} \,,  \qquad
E^{\pm\, (r)}_{i}(p) =  E^{\pm}_{p-i} \,, \qquad i = 1, \ldots, p \,,
\label{rightgensvectorrep}
\ee
respectively. The crucial point is that the broken generators 
$E^{\pm}_{p}$ of $D_{n+1}^{(2)}$ do not belong to either the ``left'' or ``right'' 
subalgebras.

The coproducts for the ``left'' generators are given by
\begin{align}
\Delta(H^{(l)}_{j}) &= H^{(l)}_{j} \otimes \id + \id \otimes H^{(l)}_{j} \,, 
\qquad\qquad\qquad\qquad\qquad\qquad\qquad\qquad j = 
1, \ldots,  n-p \,, \non \\
\Delta(E^{\pm\, (l)}_{j}) &= E^{\pm\, (l)}_{j} \otimes e^{(\eta + i \pi)
H^{(l)}_{j} - \eta H^{(l)}_{j+1}} + e^{-(\eta + i \pi)
H^{(l)}_{j} + \eta H^{(l)}_{j+1}} \otimes E^{\pm\, (l)}_{j} \,, \qquad j 
= 1, \ldots, n-p-1 \,, \non  \\
\Delta(E^{\pm\, (l)}_{n-p}) &= E^{\pm\, (l)}_{n-p} \otimes e^{\eta H^{(l)}_{n-p}} 
+ e^{-\eta H^{(l)}_{n-p} } \otimes E^{\pm\, (l)}_{n-p} \,.
\label{coproductleft}
\end{align}
These coproducts satisfy
\be
\left[  \Delta(H^{(l)}_{i}) \,,   \Delta(E^{\pm\, (l)}_{j}) \right] = \pm \alpha_{i}^{(j)} 
\Delta(E^{\pm\, (l)}_{j})\,,  
\label{DeltaHDeltaE}
\ee
and 
\begin{align}
\Omega_{ij}^{(l)}\Delta(E_i^{+\, (l)})\Delta(E_j^{-\, (l)})-\Delta(E_j^{-\, 
(l)})\Delta(E_i^{+\, (l)})\Omega_{ij}^{(l)} 
= \delta_{i,j}\frac{\sinh\left[2\eta \sum_{k=1}^{n-p}\alpha_{k}^{(j)}\Delta(H^{(l)}_k)\right]}
{\sinh(2\eta)} \,,
\label{DeltaEDelta}
\end{align} 
where $\Omega_{ij}^{(l)}$ is given by 
\be
\Omega_{ij}^{(l)}=
\begin{cases}
e^{i \pi H^{(l)}_{\text{max}(i,j)}}\otimes\id & |i-j|=1 \mbox{ and } 1 \le 
\text{min}(i,j) \le n-p-2 \\
\quad \, \quad \id\otimes\id & \mbox{ otherwise }
\end{cases}
\,.
\label{omegal}
\ee

The coproducts for the ``right'' generators are given by
\begin{align}
\Delta(H^{(r)}_{j}) &= H^{(r)}_{j} \otimes \id + \id \otimes H^{(r)}_{j} \,, 
\qquad\qquad\qquad\qquad\qquad\qquad\qquad\qquad j = 
1, \ldots,  p \,, \non \\
\Delta(E^{\pm\, (r)}_{j}) &= E^{\pm\, (r)}_{j} \otimes e^{(\eta + i \pi)
H^{(r)}_{j} - \eta H^{(r)}_{j+1}} + e^{-(\eta + i \pi)
H^{(r)}_{j} + \eta H^{(r)}_{j+1}} \otimes E^{\pm\, (r)}_{j} \,, \qquad j 
= 1, \ldots, p-1 \,, \non  \\
\Delta(E^{\pm\, (r)}_{p}) &= E^{\pm\, (r)}_{p} \otimes e^{\eta H^{(r)}_{p}}  
+ e^{-\eta H^{(r)}_{p}} \otimes E^{\pm\, (r)}_{p} \,.
\label{coproductright}
\end{align}
These coproducts satisfy
\be
\left[  \Delta(H^{(r)}_{i}) \,,   \Delta(E^{\pm\, (r)}_{j}) \right] = \pm \alpha_{i}^{(j)} 
\Delta(E^{\pm\, (r)}_{j})\,,  
\label{DeltaHDeltaEr}
\ee
and 
\begin{align}
\Omega_{ij}^{(r)}\Delta(E_i^{+\, (r)})\Delta(E_j^{-\, (r)})
-\Delta(E_j^{-\, (r)})\Delta(E_i^{+\, (r)})\Omega_{ij}^{(r)}
= \delta_{i,j}\frac{\sinh\left[2\eta \sum_{k=1}^{p}\alpha_{k}^{(j)}\Delta(H^{(r)}_k)\right]}
{\sinh(2\eta)} \,,
\label{DeltaEDeltar}
\end{align} 
where $\Omega_{ij}^{(r)}$ is given by 
\be 
\Omega_{ij}^{(r)}=
\begin{cases}
e^{i \pi H^{(r)}_{\text{max}(i,j)}}\otimes\id & |i-j|=1 \mbox{ and } 1 \le \text{min}(i,j) \le p-2  \\
\quad \, \quad \id\otimes\id & \mbox{ otherwise }
\end{cases} 
\,.
\label{omegar}
\ee

The operator $\tilde{T}^{\pm}_{i,j}(p)$ (\ref{tildeTpm}) can be 
expressed in terms of $N$-fold coproducts of these ``left'' and 
``right'' generators, similarly to the other cases considered in 
\cite{Nepomechie:2018dsn}.

% \newpage
% \clearpage

% \bibliographystyle{utphys}
% \bibliography{refs}

\providecommand{\href}[2]{#2}\begingroup\raggedright\endgroup

\end{document}